\documentclass[floatfix,twocolumn,showpacs,preprintnumbers,amsmath,amssymb,aps]{revtex4}
\usepackage{graphicx}

\begin{document}


\title{What are scientific leaders? The introduction of a normalized impact factor}


\author{George E. A. Matsas}
\address{Instituto de F\'\i sica Te\'orica, Universidade Estadual Paulista,
         Rua Pamplona 145, 01405-900, S\~ao Paulo, SP,
         Brazil}



\begin{abstract}
We define a {\em normalized impact factor} suitable to assess in a 
simple way both the strength of scientific communities and the research 
influence of individuals. We define those ones with ${\rm NIF} \geq 1$ as 
being {\em scientific leaders} since they would influence their peers at least 
as much as they are influenced by them. The NIF is distinguished because 
(a) this has a clear and universal meaning and (b) this is robust against 
self-citation expedient. We show how a single lognormal function obtained 
from a simplified version of the NIF leads to a clear ``radiography" of the 
corresponding scientific community. As an illustration, this is eventually 
applied to analyze a community derived from the list of  {\em outstanding 
referees} recognized by the {\em American Physical Society} in 2008. 
\end{abstract}


\maketitle


One of the most challenging aspects to define comprehensive scientometric 
indexes concerns the fact that this is not obvious in general how to take 
into account the field idiosyncrasies~\cite{Narinetal76}. On the other hand, 
some agreement on what scientometric guiding criteria are useful is crucial 
to allow the proper agencies to formulate general policies and optimize 
the use of financial resources. As a consequence a continuous effort 
to improve the actual parameters can be witnessed. In order to take into 
account the publishing and citation 
traditions of different areas to rank scientific journals
some authors have suggested, for instance, alternatives to 
the {\em journal impact factor} (JIF)~\cite{Cameron05}. One such example 
was proposed by Bergstrom  
who reports the development of an algorithm which captures
the percentage of the time that library users spend with a given 
journal~\cite{Bergstrom07}: ``{\em Eigenfactor Scores} and {\em Article 
Influence Scores} rank journals much as Google ranks websites". (See
also Refs.~\cite{PalacioHuertaetal04} for previous related work.)
More recently Nicolaisen and Frandsen have defined the {\em reference 
return ratio} (3R). The 3R exhibits a strong correlation with the JIF 
yet the ``3R appears to correct for citation habits, citation dynamics, 
and composition of document types"~\cite{Nicolaisenetal08}. In the same 
year Zitt and Small proposed the {\em audience factor} 
(AU)~\cite{Zittetal08} as a way to normalize the standard JIF by the 
journal field.

As part of the broad program of defining scientometric parameters 
whose interpretation is as independent as possible from the research 
field, we define here the {\em normalized impact factor} (NIF) to 
assess {\em the strength of scientific communities} 
and {\em the  influence of individual research}. The NIF is distinguished 
because (a) this has a clear and universal meaning being applicable
with equal efficiency to individuals belonging to quite distinct communities 
and (b) this is robust against self-citation expedient. For the sake of
illustration, we use the NIF to analyze a community derived from the list 
of {\em outstanding referees}  recognized by the {\em American Physical Society} 
(APS) in 2008~\cite{OR}. We compare the NIF with the $h$-index
and show that no clear correlation between them is seen indicating that
both indexes carry different pieces of information. Eventually we discuss
the NIF limitations.
\begin{figure}
\vskip 1truecm
\includegraphics[width=7.2cm]{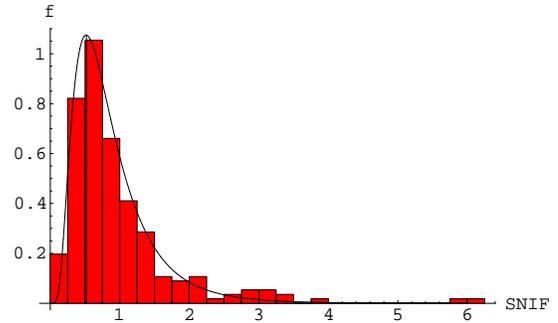}
\caption{The vertical bars show the population probability 
density obtained from our data. The area below the graph gives 
the rate of individuals with SNIF in the corresponding interval.
The solid line is a lognormal distribution fit:
$ f(x) \equiv e^{-(\ln (x)- \mu)^2/(2 \sigma^2)}/\sqrt{2 \pi~}\sigma x$
($x \equiv {\rm SNIF}$) 
with $\mu = -0.3 $ and $\sigma = 0.6$. Note that 68\% of the population 
has  $0.4  < {\rm SNIF} < 1.3$. Leaders comprise only $31\%$ of the 
population. }
\label{figfinal}
\end{figure}

\begin{figure}[!ht]
\includegraphics[width=7.2cm]{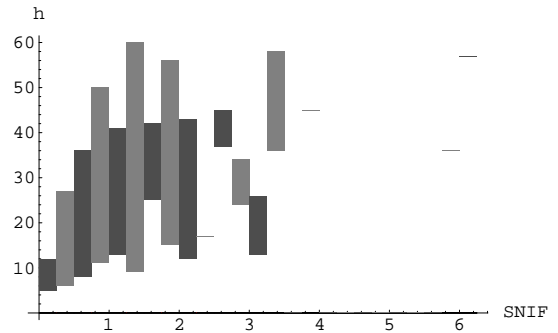}
\caption{The bars express the $h$-index range assumed by
individuals with fixed SNIF. No obvious relationship between 
the SNIF and $h$-index is seen: we can find leaders  with relatively 
small $h$-index and individuals with ${\rm SNIF}<1$ with relatively 
large $h$-index. }
\label{fit2_h}
\end{figure}

The NIF puts in context (i)~{\em the influence exerted by the research 
of an individual} with respect to (ii)~{\em how much this individual has 
been influenced by his/her scientific community}. Our assumption is that 
while {\em citations} received by an individual reflects 
(i)~\cite{Merton,PToday} the bibliographic {\em references} listed by 
him/her must reflect (ii). Hence, we define the NIF of an individual as 
\begin{equation}
{\rm NIF} \equiv \frac{\sum_i c_{i}/a_{i}}{\sum_i r_{i}/a_{i}},
\label{NIF}
\end {equation}
where $c_{i}$ and $r_{i}$ are the number of citations received 
and references included in his/her $i$th paper which is signed by 
$a_{i}$ authors, respectively. Thus, if the first paper of the 
publication list of some researcher is signed, say, by 3 authors
(the researcher himself/herself and two other collaborators),
has 30 references and has received 20 citations, then $a_1 = 3$,
$r_1 = 30$ and $c_1 = 20$, and so on for the other papers. 
The inclusion of $a_{i}$ avoids 
double counting of references and citations in multi-authored papers
leading to the following important feature: in a {\em closed community} 
of {\em identical individuals} (i.e. who publish, reference and are cited 
by each other at the same rate) all members have ${\rm NIF}=1$. This is so 
because each reference corresponds to a citation. The NIF interpretation is, 
thus, straightforward. We denominate {\em leaders} individuals with 
${\rm NIF} \geq 1$ because they influence their peers at least as much as
they are influenced by them. The existence of an universal reference value
(${\rm NIF}=1$) is one of the most satisfying aspects of this factor. 
Other ones are much field dependent and, thus, more difficult to interpret. 
Moreover, this is not easy to artificially boost one's NIF through self citation 
because the inclusion of extra references increases the denominator of 
Eq.~(\ref{NIF}). However, we note that the NIF is obviously silent about the 
intrinsic relevance of the research work itself. We might have 
leaders (in the sense that they possess ${\rm NIF} \geq 1$) influencing 
people to work on issues which eventually do not pay off their research 
effort and vice versa. The NIF is clearly unable to evaluate the 
``leadership quality".

As an illustration, we analyze a community derived from the 2008 list
of APS outstanding referees~\cite{OR}. The methodology used is as follows. 
We have randomly selected 283 individuals (among the 531 ones) and looked 
for their publication lists in the  {\em ISI Web of Science}~\cite{ISI}. 
To avoid including homonyms, we have entered with the researcher name 
{\em and} affiliation exactly as appears in Ref.~\cite{OR}, 
and recorded the ones with 20 or more papers. The  derived community
suffices for our purposes. Eventually 223 individuals (authoring 
22,611 papers) passed the criterion above, comprising 592,429 references 
and 597,390 citations. {\em Incidentally} the total number of references is 
almost the same as the total number of citations, which is a necessary 
(although not sufficient) requirement for (approximately) closed 
communities. Obviously, our 223 individuals do not form a closed community. 
Because of resource limitations we have actually carried  our analysis 
using the {\em simplified normalized impact factor} (SNIF) defined as
\begin{equation}
{\rm SNIF} \equiv \frac{\sum_i c_{i}}{\sum_i r_{i}}. 
\label{SNIF}
\end {equation}
The typical difference between the SNIF and the NIF for 10 randomly 
chosen individuals was less than 10\%. This simplified procedure has 
restricted our data collection to 3 months and final update to 4 days 
[from June, 21st to 24th (2008)]. In Fig.~\ref{figfinal} the vertical 
bars show the {\em population probability density} generated from our 
data. The area below the graph gives the rate of individuals with SNIF 
in the corresponding interval. Interestingly enough, 
Fig.~\ref{figfinal} is well fit by a lognormal distribution~\cite{lognormal}: 
$$
f(x) \equiv 
e^{-(\ln (x)- \mu)^2/(2 \sigma^2)}/\sqrt{2 \pi~}\sigma x,
$$
where $x\equiv  {\rm SNIF}$, and $\mu = -0.3 $, $\sigma = 0.6$
for the community under analysis. As a matter of fact, $\mu$ and
$\sigma$ will depend on the community but $f(x)$ is always 
unit normalized:
$$
\int_0^\infty f(x) dx = 1.
$$
We can see $\mu$ and $\sigma$ as being the ``community fingerprint".
{\em The fact that our sample comprises from high-energy experimentalists 
to condensed-matter theoreticians and our data are fit by a single 
lognormal distribution supports our hope that the (S)NIF can be successfully 
applied to quite distinct populations keeping its simple and universal
interpretation.} 
We see that 68\% of the population will be found between 
$\exp (\mu - \sigma ) \approx 0.4$ 
and
$\exp (\mu + \sigma ) \approx 1.3$.
We obtain, thus, from Fig.~\ref{figfinal} that 68\% of our population has 
$$
0.4 < {\rm SNIF} < 1.3.
$$ 
The median which separates the population in two halves with respect
to the SNIF value is given by 
$$
\widetilde {\rm SNIF} \equiv \exp(\mu ) = 0.74.
$$ 
Thus, half of the population has ${\rm SNIF} \geq 0.74$.
The SNIF expected value (or mean value) is given by 
$$
E( {\rm SNIF}) \equiv \int_0^\infty dx x f(x) = 
e^{\mu +\sigma^2/2} = 0.89.
$$ 
Note that $E( {\rm SNIF}) > \widetilde {\rm SNIF}$.
{\em It can be numerically calculated from Fig.~\ref{figfinal} that 
leaders comprise only $31\%$ of the population.} (We shall comment
that the present author does not belong to the leader set.) 

Up to this point we have used the (S)NIF to put in context the
scientific work of individuals with respect to their peers. 
Now, we go further and propose a way to assess the {\em strength} and 
{\em homogeneity} of the scientific community itself. This is possible
because of the nice fit provided by the lognormal function for the SNIF 
distribution. Let us notice that the larger the median, $\exp (\mu )$, is 
the more ``stretched" towards large NIFs Fig.~\ref{figfinal} will be.
This suggests the following definition for the {\em community strength}:
$$
{\rm CS} = \exp (\mu ).
$$ 
Similarly, the smaller the $\sigma$ is, the stiffer the graph will be,
in which case most people would have the same corresponding NIF and the
community would be considerably homogeneous. A prototype of a 
homogeneous community would be the one described in the beginning, 
namely, a closed community of identical individuals having all members 
${\rm NIF}=1$. This suggests the following relation as a measure of the 
{\em community homogeneity}:
$$
{\rm CH} = 1/\sigma.
$$ 
By the same token, the larger the $\sigma$ is the flatter the distribution 
will be, which would characterize a nonhomogeneous community consisting of 
members with quite distinct NIF values.

Now, for the sake of comparison we show in Fig.~\ref{fit2_h} the 
$h$-index~\cite{Hirsch} range (see also Ref.~\cite{Ball,Schreiber}) 
assumed by individuals in a fixed SNIF interval. No obvious 
relationship between ``large" $h$-index and scientific 
leadership in the sense defined above is seen: we find leaders with 
relatively small $h$-index and nonleaders with relatively large $h$-index. 
Notice that in general a correlation between the $h$-index and the number of
papers is expected to exist. Let us consider for a moment the extreme 
case where every papers of some individual cite all his/her previous ones. 
In this case, the corresponding $h$-index would be $N/2$ and $(N-1)/2$ for 
$N$ even and odd, respectively, only due to self citation, where $N$ is the 
total number of papers. This can lead to quite large values. Although this is 
an unrealistic case, it illustrates the possible influence of the paper number 
on the $h$-index. This is not so with the NIF which is basically insensitive
to the paper number.  The NIF is not expected to exhibit more than
a moderate correlation with the $h$-index. As a result, the pieces of 
information carried by the NIF and $h$-index are fairly distinct.

In summary, assuming that our hypotheses~(i) and~(ii) above are 
meaningful, it seems fair to support that the NIF provides useful 
and novel information about the scientific work of individuals and of
their corresponding communities. However, we would like to make some 
caution remarks concerning the use
of this index. (The same remarks apply to the SNIF.) The first one, 
is that the NIF should be used to 
assess senior rather than junior scientists. How to define seniority 
is an open question on which some agreement has to be reach. The 
NIF of junior researchers will exhibit in general a strong transient 
profile as a function of time and will be mostly influenced by the 
supervisor expertize in which case the ``leader" label would not apply. 
Moreover, the NIF does not distinguish between protagonist and 
coadjuvant authors cosigning 
the same papers. The reliability of the NIF as a way to assess 
leadership will depend on how much one's publication list 
does reflect the researcher own expertize rather than the one of 
his/her collaborators. In addition, researchers who solve a controversy 
may be less cited than the ones who have raised it. Depending on the case, 
this may induce the NIF of the former ones to be smaller than the NIF 
of the latter ones. In order to deal with these cases, other  
aspects (mostly subjective ones) should be taken into account~\cite{Zanotto}. 
Next, the NIF is 
robust against self-citation expedient but not completely immune to it. 
Self citations tend to increase and decrease the NIF value for individuals with 
${\rm NIF} < 1$ and ${\rm NIF} > 1$, respectively. If we assume the 
usual case where the {\em number of self citations} (${\rm NSC}$) is small 
in comparison with the {\em total number of citations} (${\rm TNC}$) 
received by a researcher, ${\rm NSC/TNC} \ll 1$, it is easy to get 
from Eq.~(\ref{NIF}) that 
nonleaders will have the NIF value increased due to self citation by 
a factor of order $1+ {\rm NSC/TNC} \approx 1$ having, thus, (some but) 
not a dramatic net influence on the final NIF value. The NIF can be also 
influenced depending on the individual care and generosity when 
acknowledging the influence of others by including more or less references. 
However, we do not expect it to have a huge influence either. 
Any unusual practice or artificial tentative to boost one's NIF by 
omitting crucial references should face prompt opposition from the 
peer-review system: whenever this is possible and necessary referees 
help authors to improve their papers by calling attention to related 
references. On the other 
hand, the NIF adoption would probably drive authors not to include 
irrelevant references with the aim of receiving extra citations as a 
counterpart, which we see as being a ``purifying" feature. Some concern 
can be also raised about how 
the presence of reviews in the publication list could affect the NIF. 
Although reviews tend to include long reference lists reflecting the 
contribution of various individuals to the corresponding topic, they also 
tend to be heavily cited because of its comprehensiveness. Eventually 
both tendencies should compensate each other having no major effect in 
the NIF of senior scientists. 

Despite the NIF robustness, obviously no single parameter is able 
to clear-cut leaders from nonleaders.  In spite of the possible difficulties 
which the NIF may face when applied to assess individuals, we would like 
to emphasize that our {\em main} motivation to introduce this index was 
to provide a tool to produce more accurate ``radiographies" of research 
communities as a whole (in which case individual idiosyncrasies are ``averaged"
and should not have any relevant net influence in the final diagnosis) leading 
thus to more efficient scientific policies. This 
is particularly important to communities in emergent countries which are 
still fixing their standards and aim to have eventually a protagonist role 
in the global scenario.

\begin{acknowledgments}

I am indebted to D. Gil de Oliveira and F. Montero
for help with data collection. I thank P. Ball, R. Meneghini,
J. Montero, V. Pleitez, D. Vanzella and E. Zanotto for discussions.
 Partial financial support was 
provided by Conselho Nacional de Desenvolvimento Cient\'\i fico e 
Tecnol\'ogico and Funda\c c\~ao de Amparo \`a Pesquisa do Estado
de S\~ao Paulo.

\end{acknowledgments}


\begin{thebibliography}{99}

\bibitem{Narinetal76}
G. Pinski and F. Narin, Citation influence for journal aggregates 
of scientific publications: Theory, with application to the literature 
of physics, {\em Information Processing \& Management}, 12 (1976) 297. 

\bibitem{Cameron05}
B. D. Cameron, Trends in the usage of ISI bibliometric data: Uses, 
abuses and implications, {\em Libraries and the Academy}, 5 (2005) 105.

\bibitem{Bergstrom07}
C. Bergstrom, Eigenfactor: Measuring the value and prestige of scholarly 
journals, {\em College \& Research Libraries News}, 68 (2007). Retrieved 
from www.ala.org/ ala/mgrps/divs/acrl/ publications/crlnews/2007/may/ eigenfactor.cfm

\bibitem{PalacioHuertaetal04}
I. Palacios-Huerta and O. Volij, The measurement of intellectual influence,
{\em Econometrica}, 72 (2004) 963.

\bibitem{Nicolaisenetal08}
J. Nicolaisen and T. F. Frandsen, The reference Return Ration, 
{\em Journal of Informetrics}, 2 (2008) 128.

\bibitem{Zittetal08}
M. Zitt and H. Small, Modifying the Journal Impact Factor by Fractional 
Citation Weighting: The Audience Factor, {\em Journal of the American
Society for Information Science and Technology}, 59 (2008) 1856.

\bibitem{OR}
{\em Outstanding Referees Program}, American Physical Society
(publish.aps.org/ OutstandingReferees).

\bibitem{Merton}
R. K. Merton, {\em The Sociology of Science}. 
Chicago, University of Chicago Press (1973).

\bibitem{PToday}
S. Redner,  Citation statistics from 110 years of Physical Review, 
{\em Phys. Today},  58 (2005) 49.

\bibitem{ISI}
{\em ISI Web of Knowledge}, Thomson Reuters
(www.isiwebof knowledge.com).

\bibitem{lognormal}
J. Aitchinson and J. A. C. Brown, {\em The Lognormal Distribution}. 
Cambridge, Cambridge University Press (1957).

\bibitem{Hirsch}
J. E. Hirsch, An index to quantify an individual's scientific research output, 
{\em Proc. Natl Acad. Sci. USA},  102 (2005) 16569.

\bibitem{Ball}
P. Ball, Index aims for fair ranking of scientist, 
{\em Nature}, 436 (2005) 900. 

\bibitem{Schreiber}
M. Schreiber, To share the fame in a fair way, $h_m$ modifies $h$ for 
multi-authored  manuscripts,
{\em New J. Phys.}, 10 (2008) 040201. 

\bibitem{Zanotto} 
E. D. Zanotto, The scientists pyramid, 
{\em Scientometrics}, 69 (2006) 175.

\end{thebibliography}
\end{document}